\documentstyle[psfig,12pt,aaspp4]{article}
 
\def\kms{km~s$^{-1}$}
\def\hal{H$\alpha$}
\def\be{\begin{equation}}
\def\ee{\end{equation}}
\def\about{$\sim$}

\def\HI{\ion{H}{1}}
\def\HII{\ion{H}{2}}
\def\ISOcolor{${I_\nu (6.75 \mu {\rm m})} \over {I_\nu (15 \mu {\rm m})}$}
\def\ISOcolorc{${I_\nu^{\rm c} (6.75 \mu {\rm m})} \over {I_\nu^{\rm c} (15 \mu {\rm m})}$}
\def\ISOcolora{${I_\nu^{\rm a} (6.75 \mu {\rm m})} \over {I_\nu^{\rm a} (15 \mu {\rm m})}$}
\def\IRAScolor{${I_\nu (60 \mu {\rm m})} \over {I_\nu (100 \mu {\rm m})}$}
\def\IRAScolorb{${I_\nu (12 \mu {\rm m})} \over {I_\nu (25 \mu {\rm m})}$}
\def\Nd{$N_{\rm d}$}

\begin{document}
\title{Towards an Understanding of the Mid-Infrared Surface Brightness of Normal Galaxies}
 
\author {Daniel A. Dale, George Helou, Nancy A. Silbermann, Alessandra Contursi}\affil{IPAC, California Institute of Technology 100-22, Pasadena, CA 91125}
 
\author {Sangeeta Malhotra\altaffilmark{1}}\affil{Kitt Peak National Observatory, P.O. Box 26732, Tucson, AZ 85705}
\altaffiltext{1}{Hubble Fellow}

\author {Robert H. Rubin}\affil{NASA/Ames Research Center, MS 245-3, Moffett Field, CA 94035}
 
\begin{abstract}
We report a mid-infrared color and surface brightness analysis of IC 10, NGC 1313, and NGC 6946, three of the nearby galaxies studied under the Infrared Space Observatory Key Project on Normal Galaxies.  Images with $<$9\arcsec\ (170 pc) resolution of these nearly face-on, late-type galaxies were obtained using the LW2 (6.75 $\mu$m) and LW3 (15 $\mu$m) ISOCAM filters.  Though their global \ISOcolor\ flux ratios are similar and typical of normal galaxies, they show distinct trends of this color ratio with mid-infrared surface brightness.  We find that \ISOcolor $\lesssim 1$ only occurs for regions of intense heating activity where the continuum rises at 15 $\mu$m and where PAH destruction can play an important role.  The shape of the color-surface brightness trend also appears to depend, to the second-order, on the hardness of the ionizing radiation.  We discuss these findings in the context of a two-component model for the phases of the interstellar medium and suggest that star formation intensity is largely responsible for the mid-infrared surface brightness and colors within normal galaxies, whereas differences in dust column density are the primary drivers of variations in the mid-infrared surface brightness between the disks of normal galaxies. 
\end{abstract}
 
 
\section {Introduction}
What determines the infrared surface brightness in normal galaxies?  Is the heating of interstellar dust strictly governed by young, hot stars?  How do the different populations of dust grains respond to changes in the radiation field?  What can we learn from colors?  The answers to these questions are key to understanding the large-scale physical behavior of the interstellar medium (ISM).  We pursue some of these questions here by studying mid-infrared dust emission from three nearby galaxies, seeking in particular to understand variations in mid-infrared colors and surface brightness.

The unprecented sensitivity at infrared wavelengths afforded by the Infrared Space Observatory (ISO; Kessler et al. 1996) allowed studies of the dust emission in extragalactic systems other than IR-bright starburst and ultraluminous galaxies.  The ISO Key Project on Normal Galaxies, carried out under NASA Guaranteed Time, was designed to probe the ISM in some 70 galaxies for which the luminosity is primarily derived from star formation (Helou et al. 1996).  Thanks to their proximity and favorable inclinations, IC 10, NGC 1313 and NGC 6946 were selected from the Normal Galaxy sample to be studied in detail using a variety of instruments aboard ISO.  Probing to fainter surface brightness levels than heretofore possible, ISOCAM (Cesarsky et al. 1996a) was used to construct well resolved maps of these three galaxies at 6.75 $\mu$m and 15 $\mu$m.

From elementary equations of dust heating and radiative transfer, a simple scaling for the infrared surface brightness in normal galaxies is 
$I_\nu \propto N_{\rm d} \cdot U$, where \Nd\ is the dust column density and $U$ represents the heating intensity.  In real galaxies, it is more appropriate to express the infrared surface brightness as a sum over different column densities and heating intensities along the line of sight.  Regardless of how complicated the column density and heating terms are, it is still useful to try to disentangle their relative influences on infrared surface brightness.  Helou (1986) used the IRAS color-color diagram to show that increases in the \IRAScolor\ ratio (and decreases in \IRAScolorb\ ratio) correspond to increases in star formation and heating intensity.  Unfortunately, the effective resolution of IRAS HiRes maps does not allow spatially-detailed studies of infrared surface brightness even for the largest nearby galaxies.  The resolution and sensitivity of ISO data, however, bring us closer to an answer.  For example, observations of M51 establish links between \ISOcolor, $U$, and mid-infrared surface brightness.  In addition to seeing \ISOcolor\ peak for the arms and nucleus, Sauvage et al. (1996) show the \hal\ map for this galaxy bears a one-to-one correspondence with the 15 $\mu$m image.  

In contrast to the above example, recent results for the quiescent galaxy M31 show a strong correlation between the distribution of the 7 $\mu$m, 15 $\mu$m, CO (1--0), and \HI\ emission.  However, the H$\alpha$ map shows a marked difference (Pagani et al. 1999); the \hal\ emission appears to arise from regions peripheral to the mid-infrared emission.  Moreover, the distribution of 7 $\mu$m and 15 $\mu$m emission is nearly anti-correlated with the distribution of emission at ultraviolet wavelengths.  Though the latter effect may be largely attributed to the effects of extinction, the authors show that the distribution of interstellar material is the primary driver of mid-infrared flux; the variations in the mid-infrared surface brightness for M31 are not critically dependent on the distribution of ionizing sources, but rather primarily reflect the distribution of dust.
  
A final example comes from prior work done on NGC 6946.  The mid-infrared arm/inter-arm contrast is smaller than that seen in \hal\ (Malhotra et al. 1996), and the \ISOcolor\ ratio is basically constant across the entire disk and thus over a large range in surface brightness (Helou et al. 1996).  Since mid-infrared colors are expected to vary with changing heating intensity (Helou et al. 1999b; Vigroux et al. 1998), the resulting conclusion is that surface brightness variations in the disk of NGC 6946 are largely determined by filling factor variations.  To be fair we should point out that the mid-infrared color of NGC 6946's disk is constant only to within the 9\arcsec\ resolution of ISOCAM.  As we shall see later, the mid-infrared colors of individual \HII\ regions, unresolved by ISOCAM at the distance of NGC 6946, depart significantly from the average color of the disk.

We expand upon the work of Malhotra et al. (1996) and Helou et al. (1996) by studying the mid-infrared emission of IC 10 and NGC 1313 in addition to that of NGC 6946.  We briefly discuss these galaxies and the data in Section 2.  In Section 3 we present the color-surface brightness trends for the galaxies, and in Section 4 interpret these results in the context of a simple model for the mid-infrared surface brightness of normal galaxies.  The last section summarizes our work and its implications.   

\section{The Data}
The mid-infrared imaging data used for this work are part of a much larger database we have compiled for some 70 normal galaxies.  The term ``normal'' here implies that the bulk of the luminosity should stem from stars and typical star forming regions.  A full description of the entire galaxy sample and the methods used for processing the ISOCAM data is given in Silbermann et al. (1999).   We only mention here that we employ the ``fit3'' method for transient correction since it produces, on average for the entire set of Key Project CAM obsevations, maps with relatively high signal to noise.

\subsection{ISOCAM Observations}
Mid-infrared maps of the galaxies were obtained using the LW2 (6.75 $\mu$m, $\Delta \lambda=3.5$ $\mu$m) and LW3 (15 $\mu$m, $\Delta \lambda=6.0$ $\mu$m) ISOCAM filters (note: we use the same data for NGC 6946 that were presented in Helou et al. (1996) and Malhotra et al. (1996), but we employ the more recent reduction methods described in Silbermann et al. 1999).  The observations were carried out in raster mode, with a set of 8$\times$8 pointings for NGC 6946, a set of 4$\times$4 pointings for IC 10, and two sets of 4$\times$4 pointings for NGC 1313.  IC 10 was actually mapped twice, once during ISO revolution \#457 and again in revolution \#824.  The spacing between adjacent raster pointings is 81\arcsec\ (13.5 pixels) to allow better spatial sampling -- the final maps are constructed on a grid with 3\arcsec\ pixels, though the observations were executed with the CAM pixel size set to 6\arcsec.   Table \ref{tab:cam} gives the details of the ISOCAM imaging.  The LW2 and LW3 frames have been registered and smoothed to a similar resolution (\about 9\arcsec).  

The mid-infrared band sees a transition from stellar emission to the ISM in galaxies.  In the case of these galaxies the LW2 and LW3 filters predominantly sample dust emission.  The flux seen through the LW2 filter includes strong ``aromatic features in emission'' at 6.2, 7.7, and 8.6 $\mu$m (Helou et al. 1999a).  These features are generally believed to originate from Polycyclic Aromatic Hydrocarbons (PAHs) transiently heated to high equivalent temperatures by single photons (Puget \& L\'eger 1989).  The LW3 filter, in addition to sampling the 12.5 $\mu$m aromatic feature, picks up continuum emission between 13 and 18 $\mu$m, which may be due to very small fluctuating grains, or to classical grains heated at close range by young stars.  We note that strong [NeII] (12.8 $\mu$m) and [NeIII] (15.6 $\mu$m) fine structure lines contribute significantly to the spectra of \HII\ regions (e.g. M17; Cesarsky et al. 1996b), though their overall contribution to the LW3 flux is small.  Rotational $H_2$ lines are also present in this wavelength range (Kunze et al. 1996; Valentijn et al. 1996; Timmerman et al. 1996).

\subsection{The Galaxies}
All three galaxies in our sample are well-studied at many wavelengths.  A listing of their general properties is given in Table \ref{tab:prop}.  We briefly discuss each galaxy and present their 15 $\mu$m maps with overlays of the \ISOcolor\ contours.

\subsubsection{IC 10}
Classified as a Magellanic-type dwarf irregular, IC 10 is considered to be a member of the Local Group (van den Bergh 1994); recent observations of Cepheid variables in IC 10 place the galaxy at a distance of 820 kpc (Wilson et al. 1996; Saha et al. 1996).  From the distribution and kinematics of the neutral hydrogen gas associated with this galaxy, Wilcots \& Miller (1998) show it is still in a formative stage, though there are many pockets of active star formation (0.15 $M_\odot$ yr$^{-1}$; Thronson et al. 1990).  In fact, Massey \& Armandroff (1995) find the global surface density of Wolf-Rayet stars in IC 10 is the highest in the Local Group indicating recent massive star formation.  They suggest that the numerous `holes' (of a typical size 150 pc $\times$ 100 pc) seen in the \HI\ distribution have been recently created by strong stellar winds and supernovae explosions of the massive star population that is likely associated with the Wolf-Rayet systems.

\subsubsection{NGC 1313}
Located at a distance of \about 4 Mpc (Tully 1988; de Vaucouleurs 1963), NGC 1313 is an isolated barred Sd galaxy. The rather patchy optical morphology of NGC 1313 belies a uniformly rotating \HI\ disk that is inclined at 48$^\circ$ to the line of sight (Ryder et al. 1995).  Star forming regions are evident along the bar and arms of the galaxy (Ryder et al. 1995), and there is evidence for an East-to-West star formation gradient across the galaxy (Wells et al. 1996).  This is the highest mass barred galaxy known to have no discernible radial metallicity gradient (Walsh \& Roy 1997).  Though there is no evidence at optical, radio, mid-infrared, or far-infrared wavelengths for an active galactic nucleus, X-ray data suggest an accretion-powered object ($M\gtrsim10^3M_\odot)$ \about 1 kpc north of the photometric and dynamical center (Colbert et al. 1995, Colbert \& Mushotzky 1999, and this work).

\subsubsection{NGC 6946}
NGC 6946 is a nearly face-on ($i$\about 30$^\circ$), barred Scd galaxy at a distance comparable to that of NGC 1313.  An ISOPHOT study of the far-infrared (61 $\mu$m to 205 $\mu$m) surface brightness by Tuffs et al. (1996) indicated that the bulk of the far-infrared luminosity is due to a uniformly colored and rather cold diffuse emission from the disk; a small portion of the far-infrared luminosity also arises from warmer nuclear emission.  These results agree conceptually with work done at mid-infrared wavelengths.  Outside the starburst nucleus, mid-infrared arm/interarm studies show that non-ionizing radiation is the dominant factor in heating the galaxy's dust (Malhotra et al. 1996; Helou et al. 1996).  

\section{Mid-Infrared Surface Brightness and Color Trends}

\label{sec:trends}
This paper focuses on the behavior of mid-infrared color as a function of mid-infrared surface brightness.  We use the detailed ISOCAM maps for IC 10, NGC 1313, and NGC 6946 to plot in Figure \ref{fig:color_vs_sb} the running averages of \ISOcolor\ color for different bins of mid-infrared surface brightness (the color is a ratio of surface brightnesses in units of Jy sr$^{-1}$).
The surface brightness is computed from the geometric mean of the 6.75 $\mu$m and 15 $\mu$m fluxes, using only those pixels for which the flux is at least 4 times the background rms value.  The data presented in Figure \ref{fig:color_vs_sb} for IC 10 derive from smoothed and rebinned versions of its images to simulate its appearance at a distance comparable to that of the more distant systems NGC 1313 and NGC 6946.  The 9\arcsec\ resolution corresponds to a spatial sampling scale of \about 170 pc.  Included in the plot are the arm/interarm trends for NGC 6946 given in Malhotra et al. (1996).  As expected, these trends roughly bracket our result for NGC 6946, essentially an average arm/interarm trend.  The small differences between our work and that of Malhotra et al. (1996) for NGC 6946 are attributable to the newer methods of data reduction and calibration we employ, and may be taken as a measure of the uncertainty in the data.

Both individually and collectively, the data for these three systems span a large range in surface brightness.  Though the data for the three galaxies cover different ranges of surface brightness levels, they exhibit similar trends.  In each case \ISOcolor\ is near unity over the low surface brightness ``disk'' portion of the galaxy, in agreement with measurements for Galactic cirrus (Abergel et al. 1996) and the diffuse ISM in quiescent galaxies such as NGC 7331 (Smith 1998) and M31 (Pagani et al. 1999).  Beyond a certain surface brightness level for each galaxy the color drops by a factor of 2 to 3.  The decreasing colors and increasing surface brightness levels are associated with the more active regions of the galaxies: the nucleus of NGC 6946, the arms of NGC 1313, and both the arms and nucleus of IC 10.  This is consistent with what is seen in the Antennae galaxies, where the \ISOcolor\ color ratio sharply decreases by a factor of 2.5 between the quiescent and starburst regions (Vigroux et al. 1996).  A similar phenomenon is seen in the mid-infrared spectra for the active \HII\ regions M17 (Cesarsky et al. 1996b) and N4 in the Large Magellanic Cloud (Contursi et al. 1998), though the limited spatial resolution of the CAM data prevents us from observing the \ISOcolor$\simeq$\slantfrac{1}{5} color that is typical of individual \HII\ regions.  Thus there is little doubt that \ISOcolor\ is low for regions where the radiation intensity is unusually elevated.  This data set suggests that this is a generic property of the star forming ISM in galaxies.

Additional evidence from IRAS data, which point to high heating intensities, may help to further clarify the origin of these large drifts in mid-infrared color.  The nuclear region of NGC 6946 ($r \lesssim$ 45\arcsec) responsible for the decreased \ISOcolor\ color has a \IRAScolor\ ratio of 0.98, a value which is usually associated with active and starburst galaxies (Helou 1986), as opposed to \IRAScolor\ $\approx$ 0.40 for the rest of the galaxy.\footnote{We should note that \IRAScolor \about 0.4--0.5 coincides with the demarcation between the active and cirrus components dominating the infrared emission.  As outlined in Helou (1986), the threshold is presumably not sharply delineated because the color of the active component varies from galaxy to galaxy depending on the gas density, initial mass function, and star formation history.}  In other words, \ISOcolor\ only falls below unity for the intense heating environment found in the center of NGC 6946.  The same correspondence between ISO and IRAS colors is seen in the inactive and active regions of IC 10 and NGC 1313.  

\section{The Drivers of Mid-Infrared Surface Brightness Variations}
\subsection{Physical Effects}
\label{sec:model}

Section \ref{sec:trends} describes the general similarities of the color-surface brightness trends portrayed in Figure \ref{fig:color_vs_sb}.  The differences between the curves can be explained naturally by invoking a simple scaling relation.  Surface brightness is a function of the dust column density \Nd, the dust heating intensity $U$, and the hardness or ultraviolet fraction (UVF) of the radiation responsible for the dust heating: 

\be 
I_\nu
\propto
\int U \cdot n \cdot f({\rm UVF}) \cdot {\rm d}l \;\; \sim \;\; N_{\rm d}
\cdot \left<U \cdot f({\rm UVF}) \right>.
\label{eq:model}
\ee 
Radiation hardness can be an important factor, especially near compact \HII\ regions (e.g. Cesarsky et al. 1996b).  It is admittedly difficult, though, to disentangle these parameters with observational diagnostics.  Particularly entangled are the effects of $U$ and UVF; regions exhibiting hard ultraviolet radiation, such as \HII\ regions, are generally also intense heating environments.  Ultraviolet and \hal\ photons can characterize the hardness of the radiation field, but they are significantly extincted by dust.

The entanglement between hardness and intensity of the radiation field comes 
about primarily because these two factors have similar effects on the 
mid-infrared emission.  They induce at least two physical effects which 
contribute to the drop in \ISOcolor.  Cesarsky et al. (1996b) have shown 
that in localized, very intense radiation environments mid-infrared emission features are attenuated with respect to the total infrared emission, presumably due to PAH destruction by ultraviolet photons.  This is primarily a hardness effect.  The destruction of PAHs would lead to the observed trends since PAH features dominate the LW2 flux, whereas the LW3 filter mostly samples continuum emission (Abergel et al. 1996; Boulanger et al. 1996). More recently, data from Galactic clouds and reflection nebulae have been used to argue that \ISOcolor\ depends only on the hardness of the ultraviolet radiation (Uchida et al. 1999).  On the other hand, as dust heating increases, the 15 $\mu$m continuum increases steeply compared to the continuum at 6.75 $\mu$m, reflecting a significant contribution by dust at color temperature 100 K $<$ T$_{\rm MIR}$ $<$ 200 K typical of a heating intensity up to 10$^4$ times that of the local diffuse interstellar radiation field (Helou 1998).  This is primarily an intensity effect.  Because the effective spatial resolution of the images is about 170 pc, one cannot rule out that both effects contribute to the observed behavior in most cases.  In addition to these two effects, one has to take into consideration the intermediate population of ``Very Small Grains,'' feature-less fluctuating grains which have been introduced to account for the 25 $\mu$m emission of interstellar dust (Puget \& L\'eger 1989; D\'esert, Boulanger \& Puget 1990).  This population might play a dominant role at 15 $\mu$m, yet it is hard to isolate because of a lack of spectral signature.  Because of this we combine it with the population of hot large grains for the purposes of this discussion.

To help determine qualitatively which of these effects drives the mid-infrared color variations in these disks, we turn to ISOCAM CVF (Circular Variable Filter) data taken for NGC 6946.  The observations and data reduction process for the CVF data are described elsewhere (Contursi et al., in preparation).  Displayed in Figure \ref{fig:cvf} is the average CVF spectrum for the nuclear region of NGC 6946 in the wavelength range where the Aromatic Features dominate.
The Aromatic Features are clearly present in the core, where the intensity is greatest, and the colors furthest away from the disk average.  We can thus rule out the complete destruction of PAHs by hard UV radiation as the sole reason for the mid-infrared color shift, at least within the center of NGC 6946.  Thus, the large \ISOcolor\ ratios seen for NGC 6946 are caused either by the steepening of the continuum at 15 $\mu$m for intensely heated regions, or by a combination of
that effect with PAH destruction.

Having ruled out PAH destruction as the sole driver for the colors, we now argue that pure heating of classical dust grains cannot explain the observed trends either.  In each of the galaxies represented in Figure \ref{fig:color_vs_sb}, the surface brightness increases by factors of 10 to 50 before the colors start turning over.  If the surface brightness variation was due to the increased heating of the same column of dust, this would translate into an equivalent  increase of the radiation intensity, since the Aromatic Feature emission is in the mode of fluctuations and is optically thin, so that the emerging intensity is proportional to the heating.  An increase in the heating by a factor of 10--50 would result in an increased temperature of the large grains emitting at equilibrium by a factor on the order of two, with a slight dependence on assumed optical properties (i.e. $I_\nu \propto T^5_{\rm dust}$ for a grain emissivity proportional to $\nu$).  Such a rise in temperature is not nearly sufficient to move the spectral rise towards longer wavelengths into the 15 $\mu$m range thus causing the presumed spectral steepening.  We can thus rule out a color shift driven purely by an increased heating intensity, and we are left with a picture where the \ISOcolor\ ratio is driven by the mixing of two components, one Aromatic-rich and relatively flat in the mid-infrared, and the other Aromatic-depleted due to PAH destruction by UV radiation from young stars, and with a steep spectral ascent between 7 and 15 $\mu$m.

In this picture, surface brightness reflects the combination of these two components with variable filling factors, projected within the beam and well below unity.  If the filling factors which determine surface brightness behaved as independent variables, the curves relating surface brightness and color in Figure \ref{fig:color_vs_sb} might have a random shape in any given galaxy.  However, the three disk galaxies considered here share a strikingly similar shape to that relation.  This suggests that star formation in typical disks follows a geometry that promotes this general relation.  For instance, if star forming regions are extended complexes of high column density and elevated radiation fields, \HII\ regions where \ISOcolor\ is systematically low would always be the brightest, and would be surrounded by gradually lower surface brightness emission from photo-dissociated regions at the standard \ISOcolor\ ratios.

\subsection{A Simple Model}

Even though the shape of the color-brightness relation may be dictated by geometry rather than physical processes, one can formulate a simple model based on the assumption that the shape is relatively invariant within an individual disk, and use that model to understand the variations in this relation among galaxies.  The curve representing the relation in a given disk could easily be derived from a linear combination of two components characterized by surface brightness and color: $I_\nu^{\rm c}(\lambda)$ where \ISOcolorc$\simeq$1 for the quiescent cirrus portions of the disk, and $I_\nu^{\rm a}(\lambda)$ combined with \ISOcolora$\simeq$\slantfrac{1}{5} for the brightest, actively star-forming regions.  If the contributions for the components are characterized by weighting parameters $w^c$ and $w^a$, then we have
\be
I_\nu(\lambda)=w^c I_\nu^{\rm c}(\lambda) + w^a I_\nu^{\rm a}(\lambda).
\label{eq:two_component}
\ee
The curves within the inset of Figure \ref{fig:color_vs_sb} illustrate such a linear combination model.  Such curves might extend to much lower surface brightnesses at constant cirrus color, arising in parts of the disk where no active component contributes to the emission.  The shape of a curve is determined by the range of surface brightness over which each component contributes a finite amount.

The differences between the inset curves exemplify the effects of different environments.  The short-dashed curve illustrates a curve similar to the long-dashed curve, but shifted to a higher surface brightness, as would be expected if the dust column density in the disk was scaled up while the heating intensity remained the same.  The solid curve on the other hand shows the behavior expected if there is a narrower difference in surface brightness between the quiescent and active regions.  This would be observed, for instance, in cases where the column density dropped as the heating intensity increased, partially compensating and slowing down the increase in surface brightness (see Equation \ref{eq:model}).  Finally, the dotted curve illustrates a trend where the column density and heating intensity are both increasing towards active regions faster than in the first case discussed (see Equation \ref{eq:model}).
 
\subsection{Confronting the Model With the Data}

Equation \ref{eq:model} provides a guide for the scaling behavior of these curves.  Assuming the threshold in heating intensity required to produce \ISOcolor\ colors below unity is approximately the same from galaxy to galaxy, the relative surface brightness 
levels for the inflection points of two curves are described by 
\be 
\left.{I_{\nu_1} \over I_{\nu_2}}\right|_{\rm \tiny knee} \!\!\!\!\!\!\!\sim {N_{\rm d_1} \over N_{\rm d_2}}.
\label{eq:ratio}
\ee
Do the data support this relation?  Table \ref{tab:HI} lists estimates for the metallicities and combined \HI +${2 \over 3}$H$_2$ column densities for the transition regions in these galaxies, defined by the inflection points in the color-brightness trends -- the column densities correspond to values from literature maps for the regions where the mid-infrared color falls below unity.  We have inserted a factor of two-thirds in front of the molecular hydrogen column density since molecular hydrogen has twice the mass of atomic hydrogen, but emits only one-third the infrared flux per unit mass due to the shielding effects associated with molecular clouds (Boulanger \& P\'erault 1988).  These numbers are used to compute {\it dust} column densities, scaled according to metal content.  Here we assume a dust to gas ratio of $5 \times 10^{-3}$, an approximate value for spiral galaxies  (see Mayya \& Rengarajan 1997 and references therein).  We find that the dust column density for the transition region of NGC 6946 exceeds by an order of magnitude the transition region dust column densities of IC 10 and NGC 1313.  This is consistent with the relative surface brightness levels for the three inflection points -- it could be argued that they also exhibit order of magnitude differences, in accordance with Equation \ref{eq:ratio}.  This suggests that the mid-infrared color-brightness curves for galaxies can be simply reproduced by scaling by the average dust column density.

Estimates for the threshold heating intensity in these galaxies can be made
using the infrared emissivities per hydrogen atom, calibrated by IRAS
measurements of Galactic cirrus (Boulanger \& P\'erault 1988).  In the LW2
and LW3 passbands, these numbers are $1.0 \cdot 10^{-31}$ W atom$^{-1}$ and
$0.86 \cdot 10^{-31}$ W atom$^{-1}$, respectively (B. Reach, private
communication), or alternatively, 
\be {I_\nu(6.75 \;\mu {\rm m}) \over N_{\rm HI}} = {0.018 \;{\rm MJy \;sr^{-1}} \over (10^{20} {\rm \;cm}^{-2})}; \;\;\;\;\;{I_\nu(15 \;\mu {\rm m}) \over N_{\rm HI}} = {0.035 \;{\rm MJy \;sr}^{-1} \over (10^{20} {\rm \;cm}^{-2})}.  
\ee 
Assuming Galactic cirrus regions are heated by an average interstellar flux of
$1.7G_o=2.7 \cdot 10^{-3}$ erg cm$^{-2}$ s$^{-1}$ (Draine 1978), the
proportionality constant in $I_\nu \propto N_{\rm d} U$ is of the order 3 ${\rm MJy \;sr^{-1} \;(10^{20} cm^{-2})^{-1}}\;G_o^{-1}$. Using the dust column densities and surface brightness values for the transition regions in our sample of three galaxies, we find the threshold heating intensity is approximately $U_{\rm th}=20G_o$.  This value is much smaller than the threshold heating intensities found for individual \HII\ regions, objects for which the much higher spatial resolution available allows detailed studies of the effects of spatial variations in heating intensity.  For example, Contursi et al. (1998) show that the \ISOcolor\ ratio near the N4 \HII\ region (for which CAM observations have an effective resolution of \about 2 pc) departs significantly from the cirrus value of unity only for regions where $U \gtrsim 10^3 G_o$.  A similar scenario is described by Cesarsky et al. (1996b).  The limited spatial resolution for extragalactic observations prevents us from observing phenomena associated with isolated \HII\ regions.  Instead, along any given line of sight we are likely to be witnessing contributions from several clumps of dusty material, each heated by a different radiation field; $20G_o$ is the mean $U_{\rm th}$ over \about 170 pc of star forming regions.  The superposition of flux from these clumps can lead to colors characteristic of high heating intensity at apparently moderate mid-infrared surface brightness.  This is precisely the essence of the two-component description of the emission, and augments the picture self-consistently.  

The color-surface brightness trends can also be used to compute the relative contributions of the quiet and active regions within different regions of the galaxies.  In the context of the two-component model described by Equation \ref{eq:two_component}, we can express the ordinates in Figure \ref{fig:color_vs_sb} as 
\be
{I_\nu(6.75 \mu{\rm m}) \over I_\nu(15 \mu{\rm m})} = 
{w^{\rm c} I_\nu^{\rm c}(6.75 \mu{\rm m}) + w^{\rm a} I_\nu^{\rm a}(6.75 \mu{\rm m}) \over 
 w^{\rm c} I_\nu^{\rm c}(15   \mu{\rm m}) + w^{\rm a} I_\nu^{\rm a}(15   \mu{\rm m})},
\label{eq:abscissa}
\ee 
whereas the abscissae are given by
\be
I_\nu = 
\sqrt{[w^{\rm c} I_\nu^{\rm c}(6.75 \mu{\rm m}) + w^{\rm a} I_\nu^{\rm a}(6.75 \mu{\rm m})]
\times
      [w^{\rm c} I_\nu^{\rm c}(15   \mu{\rm m}) + w^{\rm a} I_\nu^{\rm a}(15   \mu{\rm m})]}.
\label{eq:ordinate}
\ee  
If we assume $N_{\rm HI}=10^{19}$ cm$^{-2}$ and $U=1.7G_o$ for cirrus regions, then from the discussion in the preceding paragraph we infer $I_\nu^c \sim 0.5$ MJy sr$^{-1}$.  Likewise, if we assume $U=10^3 G_o$ for active regions, then $I_\nu^a \sim 3 \times 10^2$ MJy sr$^{-1}$ for the same $N_{\rm HI}$.  Equations \ref{eq:abscissa} and \ref{eq:ordinate} can be solved for $w^c$ and $w^a$; their ratio, scaled by $I_\nu^{\rm c}/I_\nu^{\rm a}$, is displayed as a function of mid-infrared surface brightness in Figure \ref{fig:w_vs_sb}.
As expected, the ratio of quiescent to active contributions is roughly constant over the disk portions of the galaxies, and the ratio decreases as the more active regions are approached.  In addition Figure \ref{fig:w_vs_sb} shows us something more telling:  the increase in total surface brightness within a disk does not result mostly from the addition of an active component.  Instead, we see that quiescent regions also increasingly contribute to the overall surface brightness as we move from low surface brightness to high surface brightness, star forming regions.  Going from the disk of NGC 6946 to its active nuclear region, the contribution of the active component increases to $w^{\rm a} I_\nu^{\rm a}$ \about 90 MJy sr$^{-1}$, while the cirrus component goes up by $w^{\rm c} I_\nu^{\rm c}$ \about 60 MJy sr$^{-1}$; the cirrus-like component goes from contributing more than 90\% of the surface brightness to about 40\% in the nuclear regions.

Interestingly, the fits that are overlayed to the curves in Figure \ref{fig:w_vs_sb} clearly show a steeper slope for IC 10 than for NGC 1313 and NGC 6946.  This can be linked to stellar content.  As mentioned earlier, IC 10 is home to a large concentration of Wolf-Rayet stars.  Such stars are associated with harder UV radiation than for OB stars since their effective surface temperatures range anywhere from 20,000 K to over 200,000 K (Langer et al. 1988).  The steep slope for the IC 10 trend could be produced in part by PAH destruction due to the hard UV radiation.  Moreover, Wolf-Rayet stars are an indication of a past-prime starburst, so a substantial fraction of the molecular clouds associated with the starburst has dissipated.  Thus the column density surrounding the active regions is lowered, which should contribute to an abrupt transition region (cf. the solid curve in the Figure \ref{fig:color_vs_sb} inset).

\section{Conclusions}

Large maps at 6.75 and 15 $\mu$m for three nearby late-type galaxies can be
interpreted in the context of a simple scaling relation for the drivers of
mid-infrared surface brightness variations in normal galaxies.  The
relation merely states that the surface brightness at infrared wavelengths is
a function of both dust column density and the intensity of the radiation
field heating the dust, whereas the color is only a function of the mixing ratio between ionized and non-ionized region emission within the beam.

What do the data tell us?  The data for IC 10, NGC 1313, and NGC 6946 are
consistent with both heating intensity and dust column density contributing to the rise in surface brightness.  Evidence for this comes from the trends of \ISOcolor\ color with mid-infrared surface brightness for the three galaxies in our sample.  The trends are similar in form:  each galaxy shows a constant disk color of approximately unity for a range of surface brightness levels.  Then, at a certain surface brightness level that scales according to a galaxy's dust content, the trend exhibits a characteristic decline to smaller colors that correspond to the more active regions of the galaxy.  Spectral data for NGC 6946 do not conclusively show whether these colors reflect rising continua at 15 $\mu$m or whether the destruction of PAHs plays an important role; it is likely that the two effects combined yield the observed colors.  We note that the inflections in the trends may coincide with a threshold value in heating intensity averaged over the \about 170 pc beam, as they are found to occur at the boundaries of the more active star forming regions within the galaxies.  The inferred heating thresholds are biased low, however, due to the superposition of dust emission along the line-of-sight from multiple regions, with each region home to a different radiation field.  Finally, the steepness of the downturn appears to be regulated by the hardness of the heating spectrum.  For example, the star forming irregular galaxy IC 10 is presumably a rich source of hard ultraviolet radiation, and it exhibits a relatively rapid decrease in its mid-infrared color--surface brightness trend.  This interpretation requires, however, a more detailed analysis of this effect, preferably one that profits from a direct measure of the relative UV hardness of the radiation fields in IC 10, NGC 1313, and NGC 6946.

We establish that the color-surface brightness trends are consistent with a two-component model for the phases of the ISM, one Aromatic-rich and relatively flat in the mid-infrared (the ``quiescent'' phase), and the other Aromatic-poor with a steep spectral ascent between 7 and 15 $\mu$m (the ``active'' phase); the trends are best described by a mixing of a hot active component of the ISM in conjunction with a cirrus-like component.  In this context, the data show that the surface brightness contributions of both the quiescent and the active phases of the ISM increase significantly in the directions of the high surface brightness, star forming regions.  It is somewhat of a surprise that the surface brightness of regions usually labelled ``active'' receives important contributions from the cirrus-like portions of the ISM.

The results described here lead us to argue the following important conclusion.  The primary driver of surface brightness levels {\it within} normal galaxies is the intensity of the radiation field determined by star formation activity, while differences in surface brightness {\it between} galaxy disks are largely driven by variations in dust column density. 

\acknowledgements 

This work was supported by ISO data analysis funding from the US National
Aeronautics and Space Administration, and carried out at the Infrared
Processing and Analysis Center and the Jet Propulsion Laboratory of the
California Institute of Technology.  ISO is an ESA project with instruments
funded by ESA member states (especially the PI countries: France, Germany,
the Netherlands, and the United Kingdom), and with the participation of
ISAS and NASA.

\begin{figure}[!ht]
\caption[]
{\ The contours of the IC 10 15 $\mu$m map overlayed on the \ISOcolor\ image smoothed to an overall resolution of \about 9\arcsec.  Contour separations indicate surface brightness steps of 9 MJy sr$^{-1}$; the lowest contour is at 1 MJy sr$^{-1}$.  Coordinates are given in the epoch J2000.}
\label{fig:IC10}
\end{figure}
\begin{figure}[!ht]
\caption[]
{\ Similar to Figure 1, but for NGC 1313.  Contour separations indicate surface brightness steps of 0.9 MJy sr$^{-1}$; the lowest contour is at 0.6 MJy sr$^{-1}$.}
\label{fig:NGC1313}
\end{figure}
\begin{figure}[!ht]
\caption[]
{\ Similar to Figure 1, but for NGC 6946.  Contour separations indicate surface brightness steps of 2 MJy sr$^{-1}$; the lowest contour is at 1 MJy sr$^{-1}$.}
\label{fig:NGC6946}
\end{figure}
\begin{figure}[!ht]
\caption[]
{\ The average \ISOcolor\ color ratio as a function of mid-infrared surface brightness, with the data for IC 10 simulated as they would appear at the distance of NGC 1313 and NGC 6946 (about 4 Mpc).  The model curves drawn in the inset are discussed in Section \ref{sec:model}.  We have included the arm/interarm trends from the disk of NGC 6946 (Malhotra et al. 1996).}
\label{fig:color_vs_sb}
\end{figure}
\begin{figure}[!ht]
\caption[]
{\ Mid-infrared spectrum for the inner $18\arcsec \times 18\arcsec$ of NGC 6946.}
\label{fig:cvf}
\end{figure}
\begin{figure}[!ht]
\caption[]
{\ The ratio of the weights given to the active and cirrus-like components, as a function of mid-infrared surface brightness.}
\label{fig:w_vs_sb}
\end{figure}
\begin{deluxetable}{lcccrccc}
\small
\tablewidth{7.0in}
\tablenum{1}
\tablecaption{CAM Observations}
\def\x{$\times$}
\tablehead{
\colhead{Galaxy}& \colhead{R.A.~~~~~~~~~Decl.} & \colhead{Date}& \colhead{Rev}&
\colhead{Exposure\tablenotemark{a}}&
\colhead{Raster}& \colhead{Raster Map}& \colhead{backgr. rms}\nl
\colhead{}& \colhead{(J2000)} & \colhead{}& \colhead{}&
\colhead{(seconds)}&
\colhead{Pattern}& \colhead{Size\tablenotemark{b}}& \colhead{{\footnotesize ($\mu$Jy arcsec$^{-2}$)}}\nl
\colhead{}& \colhead{} & \colhead{}& \colhead{}&
\colhead{{\footnotesize LW2~~LW3}}&
\colhead{}& \colhead{}& \colhead{{\footnotesize LW2~~LW3}}
}
\startdata
IC 10     &002024.5 $+$591730& 15 Feb 1997 & 457 &  806~~~672 & ~~~4\x4 & ~7\farcm25\x~7\farcm25 & 3~~~~~~8\nl
          &                  & 16 Feb 1998 & 824 &  806~~~840 & ~~~4\x4 & ~7\farcm25\x ~7\farcm25 & 3~~~~~~9\nl
NGC 1313  &031815.4 $-$662951& 18 Apr 1997 & 663 &  806~~~672 & 2\x4\x4 & 11\farcm85\x ~9\farcm20 & 2~~~~~~3\nl
NGC 6946  &203452.3 $+$600914& 08 Feb 1996 & 083 & 2580~~1344 & ~~~8\x8 & 12\farcm65\x 12\farcm65 & 3~~~~~10\nl
\enddata
\tablenotetext{a}{Integration time in seconds, summed over all raster positions.  At each position we took several 5 second LW2 frames and several 2 second LW3 frames.}
\tablenotetext{b}{The revolution 824 map for IC 10 was offset from the revolution 457 map by ($\Delta$RA, $\Delta$Dec)=($-$57\arcsec,$+$53\arcsec).  The two maps are registered and coadded to produce the final IC 10 map.}
\label{tab:cam}
\normalsize
\end{deluxetable}
\begin{deluxetable}{lccccccccc}
\small
\tablenum{2}
\tablecaption{Galaxy Global Properties}
\def\e{($10^{-12}$ ${{\rm W} \over {\rm m}^2}$)}
\def\d{($L_\odot$)}
\def\x{$\times$}
\tablehead{
\colhead{Galaxy} & \colhead{Type} & \colhead{$a$\x$b$} & \colhead{Distance\tablenotemark{a}} & 
\colhead{log${L_B \over L_\odot}$\tablenotemark{b}} & 
\colhead{log${M_{\rm HI} \over M_\odot}$\tablenotemark{c}} &
\colhead{log${L_{\rm FIR} \over L_\odot}$\tablenotemark{d}} & 
\colhead{$f_\nu(6.75 \mu {\rm m}) \over f_\nu(15 \mu {\rm m})$}  & 
\colhead{$f_\nu(60 \mu {\rm m}) \over f_\nu(100 \mu {\rm m})$} \nl 
\colhead{} & \colhead{} & \colhead{} & \colhead{(Mpc)} & \colhead{} &
\colhead{} & \colhead{} & \colhead{} & \colhead{}}
\startdata
IC 10    & IBm  & ~6\farcm3\x5\farcm1 & 0.8 &~~8.6 & 7.7 & 8.1 & 0.71 & 0.63\nl
NGC 1313 & SBd  & ~9\farcm1\x6\farcm9 & 3.6 &~~9.5 & 9.1 & 9.0 & 0.82 & 0.47\nl
NGC 6946 & SBcd & 11\farcm5\x9\farcm8 & 4.5 &~10.1 & 9.6 & 9.8 & 0.97 & 0.46\nl
\enddata
\tablenotetext{a}{The distance for IC 10 comes from Cepheid measurements (Saha et al. 1996).  To compute distances for NGC 1313 and NGC 6946, we remove the Virgocentric motion of the Local Group from their heliocentric redshift velocities and assume a Hubble constant of 75 \kms\ Mpc$^{-1}$.}
\tablenotetext{b}{Computed from the RC3 $B_T$ values (de Vaucouleurs et al. 1991) after accounting for internal and Galactic extinction (Schlegel, Finkbeiner, \& Davis 1998) and the RC3 $k$-correction. Assumes $F_B=\nu I_\nu$ at 4400 $\AA$ and $m_B=0$ corresponds to 4260 Jy.}
\tablenotetext{c}{Derived from the distances in column 4 and the \HI\ masses given in Shostak \& Skillman (1989; IC 10), Ryder et al. (1995; NGC 1313), and Boulanger \& Viallefond (1992; NGC 6946).}
\tablenotetext{d}{Derived from the IRAS 60 and 100 $\mu$m fluxes (Helou et al. 1988).}
\label{tab:prop}
\normalsize
\end{deluxetable}
\begin{deluxetable}{lccccr}
\small
\tablewidth{6.7in}
\tablenum{3}
\tablecaption{Transition Region Data}
\tablehead{
\colhead{Galaxy} & \colhead{$12+\log{{\rm N(O)} \over {\rm N(H)}}$} &
\colhead{$N_{\rm HI}+{2 \over 3} N_{\rm H_2}$\tablenotemark{a}} & \colhead{$N_{\rm d}$} & \colhead{ref.}\nl 
\colhead{} & \colhead{} & \colhead{($10^{20}$ cm$^{-2}$)} &
\colhead{($10^{-3} M_\odot$ pc$^{-2}$)} & \colhead{}}
\startdata
IC 10    & 8.2  & 25 & ~14 & 1,2,3,4\nl
NGC 1313 & 8.4  & 10 & ~~9 & 5,6,7\nl
NGC 6946 & 9.3  & 30 & 220 & 8,9,10,11\nl
\enddata
\tablerefs{(1) Wilcots \& Miller 1998; (2) Shostak \& Skillman 1989; (3) Lequeux et al. 1979; (4) Ohta et al. 1988; (5) Ryder et al. 1995; (6) Walsh \& Roy 1997; (7) Harnett et al. 1991; (8) Boulanger \& Viallefond 1992; (9) Pilyugin \& Ferrini 1998; (10) Tacconi \& Young 1986; (11) Belley \& Roy 1992}
\tablenotetext{a}{We use a CO to H$_2$ conversion factor of $N_{\rm H_2}/I_{\rm CO} \sim 3 \cdot 10^{20}$ cm$^{-2}$ (K km s$^{-1})^{-1}$}
\label{tab:HI}
\end{deluxetable}
\normalsize
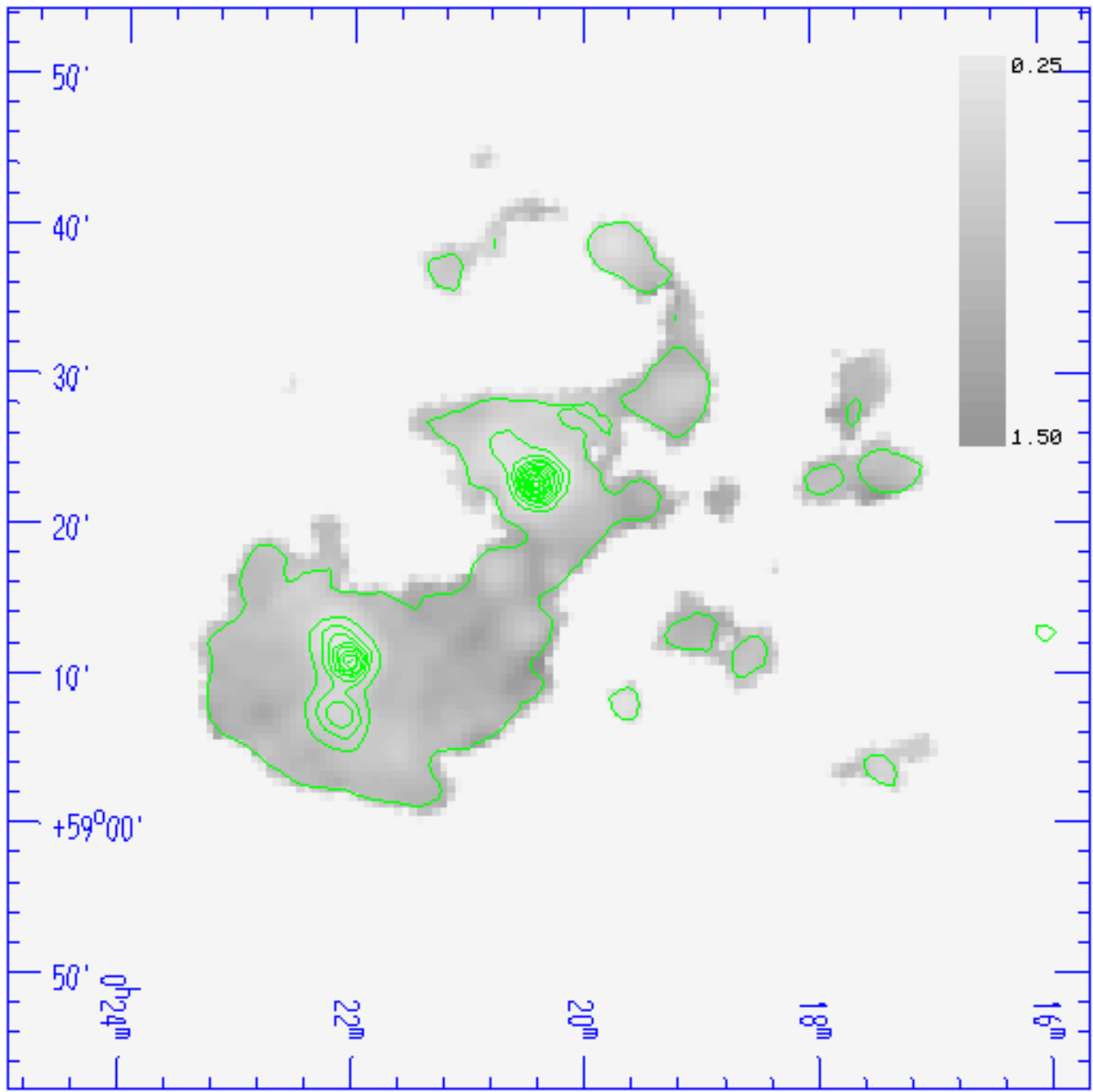
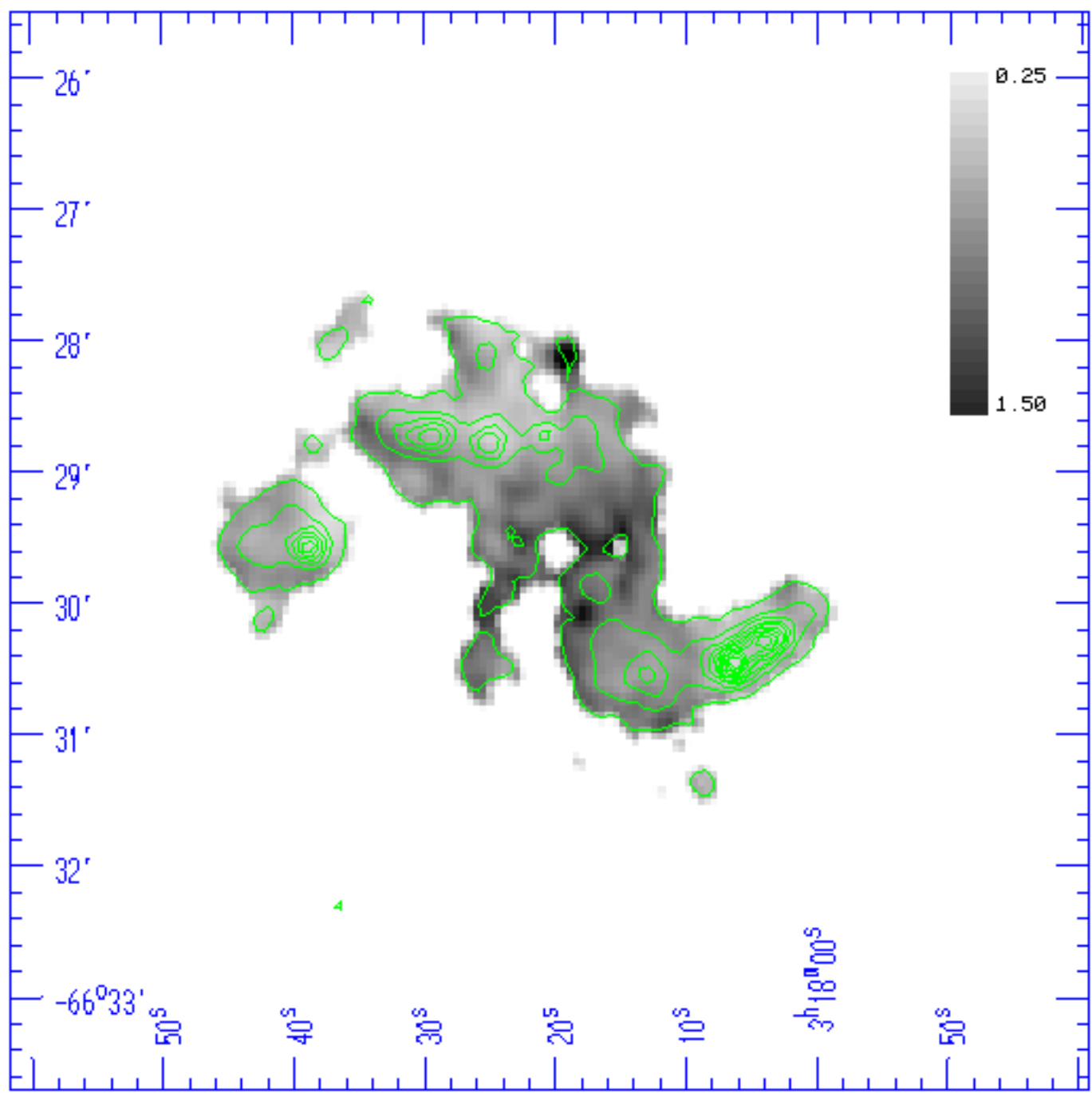
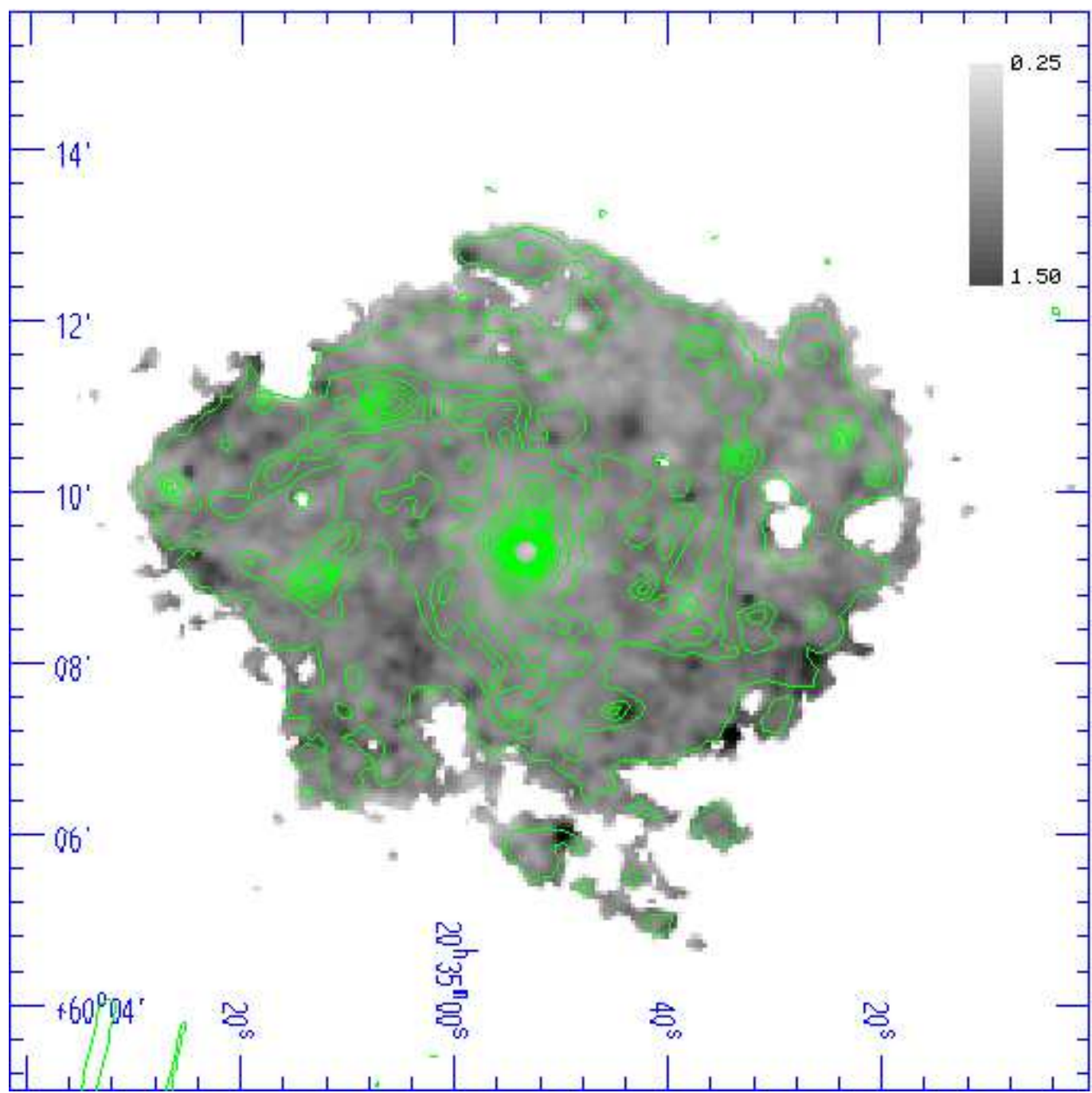
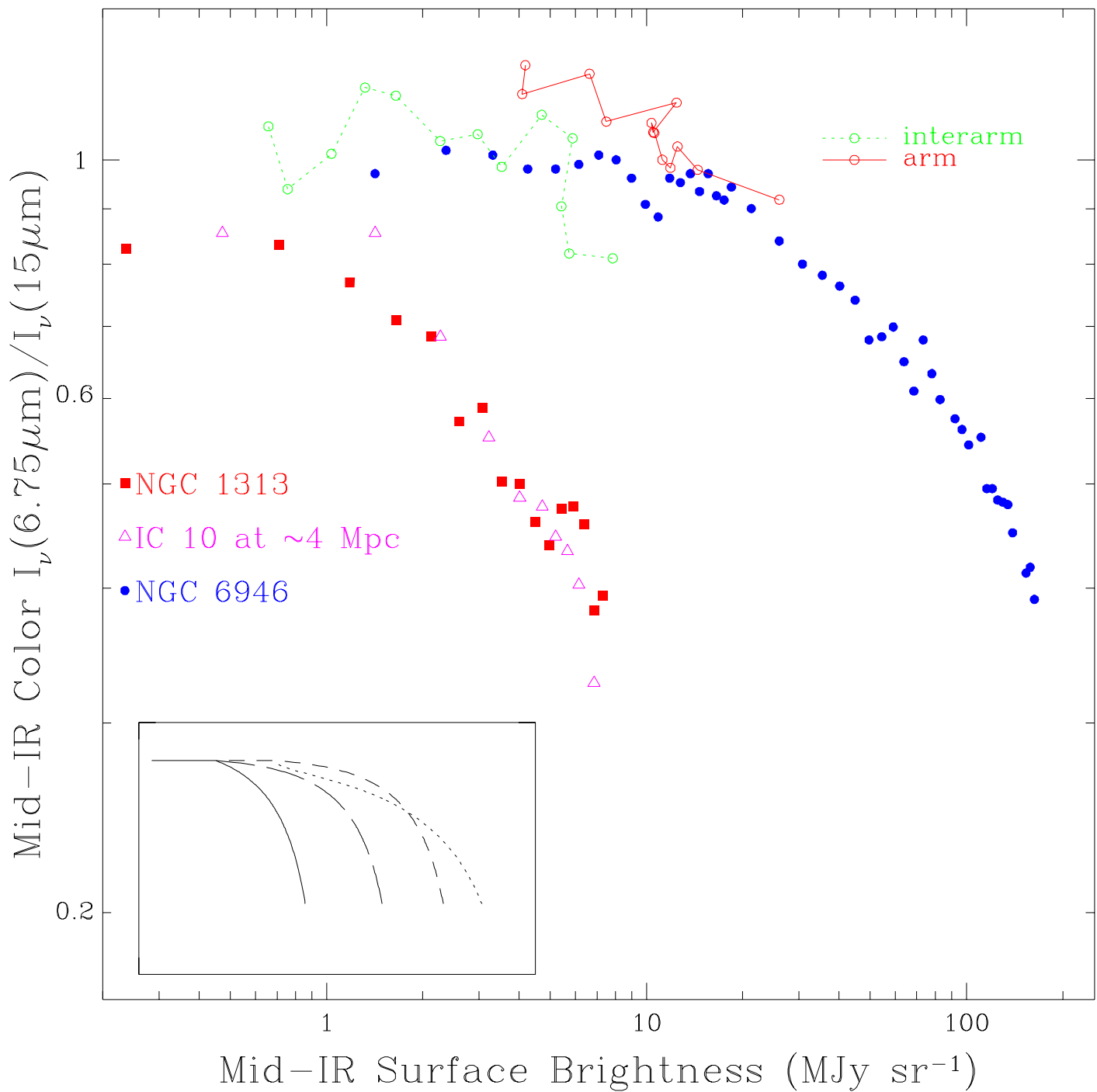
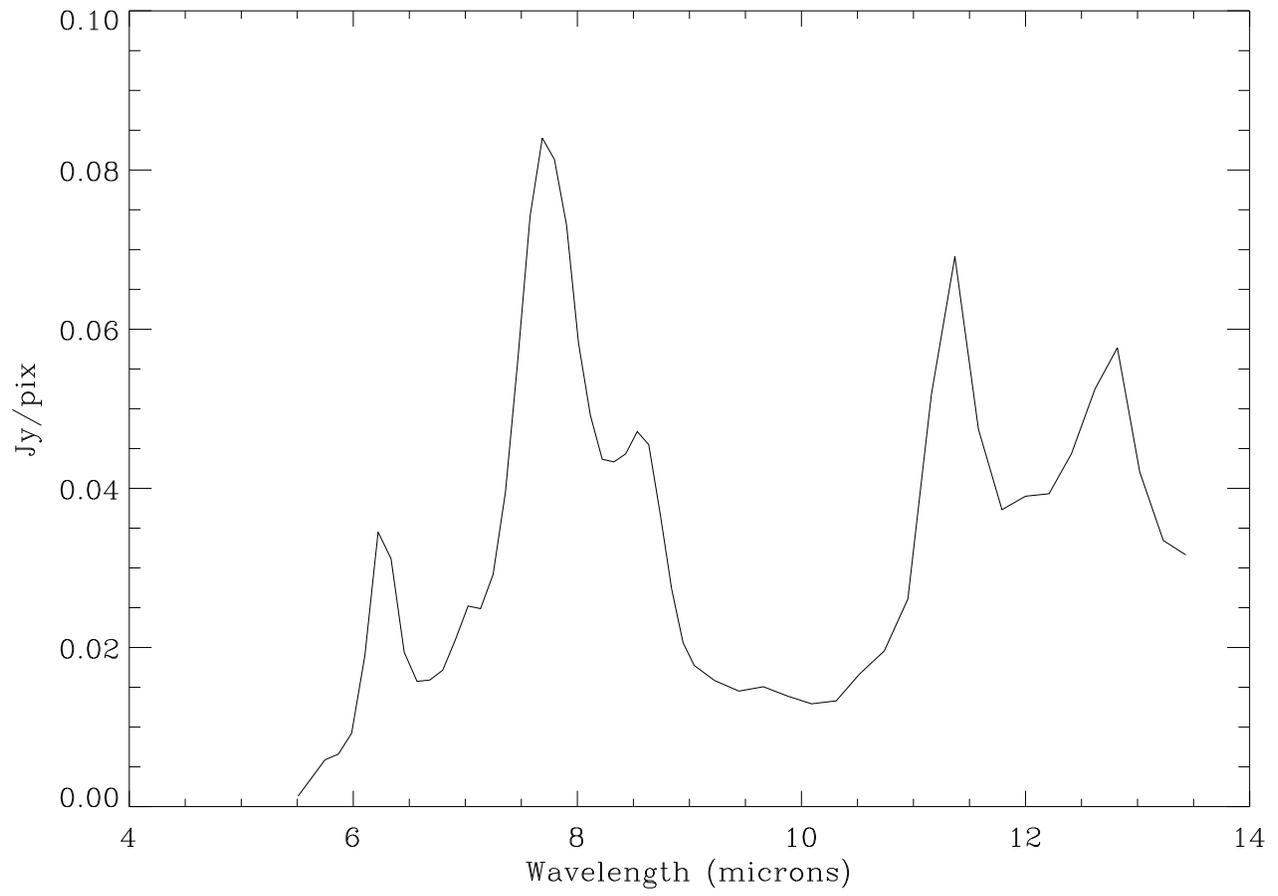
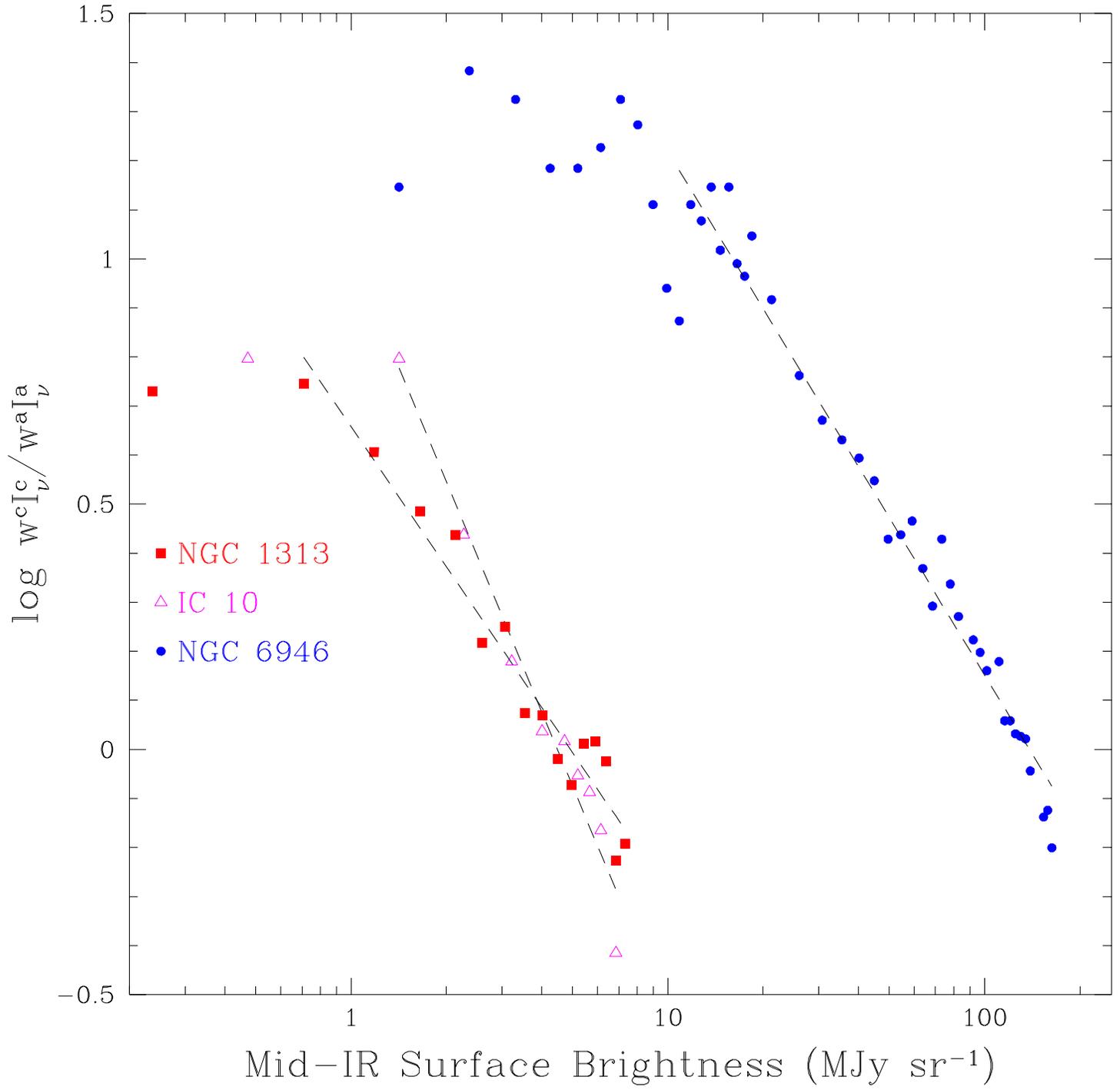
\end{document}